\documentclass[preprint,prb]{revtex4}
\usepackage{amssymb,amsmath}
\usepackage{graphicx}
\usepackage{epsfig}
\usepackage{amsmath}
\usepackage{amsfonts}

\begin{document}

\title{An anomalous particle-exchange mechanism for two isolated Bose gases
merged into one}
\author{Q. H. Liu}
\affiliation{School for Theoretical Physics, School of Physics and Electronics, Hunan
University, Changsha, 410082, China.}

\begin{abstract}
In an isolated ideal Bose system with a fixed energy, the number of
microstates depends solely on the configurations of bosons in excited
states, implying zero entropy for particles in the ground state. When two
such systems merge, the resulting entropy is less than the sum of the
individual entropies. This entropy decrease is numerically shown to arise
from an effectively yet anomalous exchange of particles in excited states,
where $\overline{N}!/(\overline{N}_{1}!\overline{N}_{2}!)<1$. Here, $%
\overline{N}$, $\overline{N}_{1}$, and $\overline{N}_{2}$ are real decimals
representing, respectively, the mean number of particles in excited states
in the merged system and the two individual systems before merging, with $%
\overline{N}<\overline{N}_{1}+\overline{N}_{2}$.
\end{abstract}

\maketitle

\section{Introduction}

The entropy of a Bose condensate is known to be zero due to quantum
coherence, allowing it to be approximated as a single macroscopic molecule
in the thermodynamic limit. \cite{BOOK1,BOOK2,BOOK3} More generally, all
particles in the ground state possess zero entropy, regardless of system
size. \cite{BEC1,BEC2,BEC3} This raises a question: how does entropy change
when two Bose systems mix or merge? While indistinguishability of particles
suggests no entropy change, the redistribution of excited-state particles
upon merging implies a change because the number of excited-state particles
in merged system is not the summation of those before mixing and the
ground-state particles are always of zero-entropy irrespective of the number
of particles. This previously unrecognized paradox is addressed by
explicitly examining ideal bosons on two-dimensional spherical surfaces
within the microcanonical ensemble. We identify a novel particle-exchange
mechanism for excited-state particles when two systems merge into one.
Although ensembles are generally considered equivalent in the thermodynamic
limit, this is not always true. Certain quantities that diverge in the
grand-canonical ensemble vanish uniformly in the microcanonical ensemble,
which is thermodynamically correct. \cite{PRL1} Furthermore, Bose-Einstein
condensation on a sphere's surface has recently been shown to be
experimentally feasible. \cite{PRL2}

In section II, ideal boson gas on the spherical surface is introduced and
what independence of the number of microstates on ground-state particle is
demonstrated. In section III, the merging of two boson gases is defined and
studied in detail, and since only the excited state particles have
entropies, a mechanism for the merging entropy change is proposed and
numerically demonstrated to be correct. This paper is enclosed by a brief
conclusions and discussions in section IV. In whole paper, $\overline{f}$
stands for ensemble mean for quantity $f$.

\section{Bosons on the spherical surface}

The energy of a particle of mass $m$ moving on the surface of a sphere of
radius $R$ is quantized and the levels are given by 
\begin{equation}
\epsilon _{l}={\frac{\hbar ^{2}}{2mR^{2}}}l(l+1),  \label{energyquantum}
\end{equation}%
where $\hbar $ is the reduced Planck's constant and $l=0,1,2,...$ is the
angular momentum quantum number and can also be called the $l$-th energy
shell. Only the energy levels $(0,2,6,12,20,30,42,...,l(l+1),...)$ $\hbar
^{2}/\left( 2mR^{2}\right) $ are possibly to accommodate particles. Once
there are $N$ noninteracting bosons moving on the surface of a sphere and
the system has definite energy ${E}$, the following partition condition must
be imposed, 
\begin{equation}
E=\sum_{l=1}^{L}n_{l}\epsilon _{l}\text{, or, }\widetilde{E}\equiv \frac{%
2mR^{2}}{\hbar ^{2}}E=\sum_{l=1}^{L}n_{l}l(l+1)  \label{TE}
\end{equation}%
where $n_{l}$ is the number of particles in the $l$-th shell and $\widetilde{%
E}$ stands for the reduced energy of the system and $L\equiv l_{\max }$
stands for the highest possible shell occupied. Evidently, we have with $%
N_{0}$ and $N_{ex}$ being respectively the number of particles in the ground
state $\epsilon _{0\text{ }}$and excited state $\epsilon _{l}$ $\left( l\geq
1\right) $, 
\begin{equation}
N=N_{0}+\sum_{l=1}^{L}n_{l},N_{ex}\equiv \sum_{l=1}^{L}n_{l}.  \label{totalN}
\end{equation}%
Given a total permissible energy $E$ or $\widetilde{E}$, we have many
different configurations and every configuration takes the following form, 
\begin{equation*}
\left\{ \mathbf{n}_{l}\right\} \equiv \{n_{L},n_{L-1},...,n_{2},n_{1}\}.
\end{equation*}%
The total number of the configurations is the number of microstates $\Omega $%
, and the entropy of the system is given by,%
\begin{equation}
S=k\ln \Omega  \label{entropy}
\end{equation}%
where $k$ is the Boltzmann constant. Though there are arbitrary number of
bosons on the sphere, the total reduced energy of the system can only take
values of zero and positive even numbers, i.e., $\widetilde{E}=$ $%
0,2,4,...,l(l+1),...$.

Let us discuss the general structure of $\left\{ \mathbf{n}_{l}\right\} $.
Once the only $L$-th shell is occupied by one particle with rest in the
ground state, we have $\left\{ \mathbf{n}_{l}\right\} =$ $\{1,0,...,0,0\}$;
and once only the first excited state are occupied with the rest in the
ground state, we have $\left\{ \mathbf{n}_{l}\right\} =$ $%
\{0,0,...,0,L(L+1)/2\}$; and other situations take the similar form in
between compatible with the constraint (\ref{TE}). Thus, there are must a
critical number of particles $N=N_{c}=L(L+1)/2$ beyond which all additional
number of particles can only be in the ground state. This system can be
called the saturated system. System with energy $\widetilde{E}=L(L+1)$ and
number of particle $N\geq N_{c}$ can thus be called supersaturated, for
further addition of particles will all increase the number of particles in
ground state with the number of microstates remaining unchanged. It is
similar to the formation of usual Bose condensate but our classification is
universal, irrespective of whether a real phase transition occurs or not.

We now illustrate in detail how to count the number of microstates. Assume
that the system has a fixed reduced energy $\widetilde{E}=12$, the energy
level of the $3$rd shell, let us consider the system has $N=1,2,3,4,5,6,7$
particles, respectively. Every configuration takes the form $\left\{ \mathbf{%
n}_{l}\right\} \equiv \{n_{3},n_{2},n_{1}\}$. We have only one configuration 
$\{1,0,0\}$ ($N_{0}=0$) for $N=1$; and two configurations $\{1,0,0\}$ ($%
N_{0}=1$)\ and $\{0,2,0\}$ ($N_{0}=0$) for $N=2$; and two configurations $%
\{1,0,0\}$ ($N_{0}=2$)\ and $\{0,2,0\}$ ($N_{0}=1$) for $N=3$; and three
configurations $\{1,0,0\}$ ($N_{0}=3$), $\{0,2,0\}$ ($N_{0}=2$), and $%
\{0,1,3\}$ ($N_{0}=0$) for $N=4$; and three configurations $\{1,0,0\}$ ($%
N_{0}=4$), $\{0,2,0\}$ ($N_{0}=3$) and $\{0,1,3\}$ ($N_{0}=1$) for $N=5$;
and four configurations $\{1,0,0\}$ ($N_{0}=5$), $\{0,2,0\}$ ($N_{0}=4$) and 
$\{0,1,3\}$ ($N_{0}=2$) $\{0,0,6\}$ ($N_{0}=0$) for $N=6$; and four
configurations $\{1,0,0\}$ ($N_{0}=6$), $\{0,2,0\}$ ($N_{0}=5$) and $%
\{0,1,3\}$ ($N_{0}=3$) $\{0,0,6\}$ ($N_{0}=1$) for $N=7$. When the number of
particles is more than $6$, there is no new configuration possible otherwise
the energy constraint is broken. We summarize the results in the following
table I 
\begin{table}[tbp]
\caption{For system $\protect\widetilde{E}=12$ and different numbers of
particles, we have different mean number of particle $\overline{N_{0}}$ in
the ground state and also its fluctuations $\overline{\Delta N_{0}}$.}%
\begin{tabular}{|c|c|c|c|c|c|c|c|}
\hline
$N$ & $1$ & $2$ & $3$ & $4$ & $5$ & $6$ & $7$ \\ \hline
$\Omega $ & $1$ & $2$ & $2$ & $3$ & $3$ & $4$ & $4$ \\ \hline
$\overline{N_{0}}$ & $0$ & $0.5$ & $1.5$ & $1.67$ & $2.67$ & $2.75$ & $3.75$
\\ \hline
$\overline{\Delta N_{0}}$ & $0$ & $0.5$ & $0.5$ & $1.25$ & $1.25$ & $1.92$ & 
$1.92$ \\ \hline
\end{tabular}%
\end{table}

\section{An entropy analysis for two isolated Bose gases merged into one}

Initially, two boson systems are separated by an impermeable spherical
membrane of fixed radius $R$, each has an energy and a certain number of
particles, $\{\widetilde{E}_{in},N_{in}\}$ and $\{\widetilde{E}%
_{out},N_{out}\}$, hereafter, the superscripts $"in"$ and $"out"$ denote the
inner and outer side of the surface, respectively. Finally, let the
impermeable membrane be permeable for both particle and energy and we have a
merged system of energy and number of particles $\{\widetilde{E},N\}=\{%
\widetilde{E}_{in}+\widetilde{E}_{out},N_{in}+N_{out}\}$. When the merging
process is adiabatic and no work done to the system, the entropy change%
\begin{equation}
\Delta S=S-\left( S_{in}+S_{out}\right) 
\end{equation}%
can be taken for granted to be zero. However, a moment of reflection gives
that it can never be true. Because number of ground-state particles in the
merged system may be more than those in two not-yet-merged systems, and
neither of them has entropy, we must have $\Delta S\leq 0$. On the other
hand, if so, the second law of thermodynamics is violated for the merging
process is entropy-invariant. We have now a paradox, and something important
is missing.

\subsection{An anomalous particle-exchange mechanism}

The particles before merging can be identified by their positions, inner or
outer of the surface. The total number of microstates after merging is
shifted to%
\begin{equation}
\Omega =\Omega _{mer}\Omega _{in}\Omega _{out}
\end{equation}%
where $\Omega _{mer}$ may take the following form, \cite{penrose} due the
particle-exchange of two distinguishable clouds of bosons,%
\begin{equation}
\Omega _{mer}=\frac{N!}{N_{in}!N_{out}!}.
\end{equation}%
However, it can never be true because effective particle-exchange can only
happen between those in excited states. Thus, we reach our key finding that
there is an anomalous particle-exchange mechanism that is defined by the
multiplication factor $\Omega _{mer}$ 
\begin{equation}
\Omega _{mer}\approx \frac{\overline{N_{ex}}!}{\overline{N_{in\text{ }ex}}!%
\overline{N_{out\text{ }ex}}!}  \label{penrose}
\end{equation}%
which contributes a new term of entropy due to merging%
\begin{equation}
S_{mer}=k\ln \Omega _{mer}
\end{equation}%
such that 
\begin{equation}
S=S_{in}+S_{out}+S_{mer}.  \label{totalS}
\end{equation}

\subsection{A numerical verification of the mechanism}

It is clearly that the relation (\ref{penrose}) can only hold approximately.
To see how accurate it is, we introduce two parameters in $\Omega _{mer}$ (%
\ref{penrose}) and the exact form of it is given by

\begin{equation}
\Omega _{mer}=\frac{\left( \overline{N_{ex}}+\mu +\nu \right) !}{\left( 
\overline{N_{in\text{ }ex}}-\mu \right) !\left( \overline{N_{out\text{ }ex}}%
-\nu \right) !}.
\end{equation}%
If our mechanism is correct, the correction parameters $\mu $ and $\nu $
must be small in comparison of either the number of particles in excited
states ($\mu \ll \overline{N_{in\text{ }ex}}=N_{in}-\overline{N_{0}}_{in}$
and $\nu \ll \overline{N_{out\text{ }ex}}=N_{out}-\overline{N_{0}}_{out}$),
or the fluctuation number of particles in excited states ($\mu \ll \overline{%
\Delta N_{in}}_{ex}=\overline{\Delta N_{0}}_{in}$ and $\nu \ll \overline{%
\Delta N_{out}}_{ex}=\overline{\Delta N_{0}}_{out}$), respectively.

We consider $\widetilde{E}=552$ only, and the saturated system has number of
particle $N_{c}=552/2=276$.

Let us first consider unsaturated systems: $N=200$ for merged systems, which
has entropy $S=15.39k$, the mean number of particles in ground state is $%
\overline{N_{0}}=143.27$ and the fluctuation is $\overline{\Delta N_{0}}%
=30.02$, We deal with two not-yet-merged systems $N_{in}=120$, $\widetilde{E}%
_{in}=342$, and $N_{out}=80$, $\widetilde{E}_{in}=210$. A large set of
of parameters $\left\{ \mu ,\nu \right\} $ solves the Eq. (\ref{totalS}),
however, we pick up the one set which makes $\left\vert \mu \right\vert
+\left\vert \nu \right\vert $ minimum. The results are listed in the table
II. The exact numerical results confirm our mechanism. 
\begin{table}[tbp]
\caption{A numerical verification of the mechanism from unsaturated systems: 
$\protect\mu \ll \min \left\{ \overline{\Delta N_{0in}},\overline{N_{in\text{
}ex}}\right\} ,\protect\nu =0$.}$%
\begin{array}{|c|c|c|c|c|c|}
\hline
\left\{ N_{in},N_{out}\right\}  & \left\{ \overline{N_{0in}},\overline{%
N_{0out}}\right\}  & \left\{ \overline{\Delta N_{0in}},\overline{\Delta
N_{0out}}\right\}  & \left\{ S_{in},S_{out}\right\} /k & \Delta S/k & 
\left\{ \mu ,\nu \right\}  \\ \hline
\{120,80\} & \{79.57,51.30\} & \{20.89,14.68\} & \{12.08,9.30\} & -5.99 & 
\{0.11,0\} \\ \hline
\end{array}%
$%
\end{table}

Now we study the saturated systems. To note that $\{\widetilde{E}_{in},%
\widetilde{E}_{out}\}$ can be $\{462,90\}$, $\{420,132\}$, and $\{342,210\}$%
, respectively, thus the numbers of particles $\{N_{in},N_{out}\}$ are $%
\{462,90\}/2=\{231,45\}$, $\{420,132\}/2=\{210,66\}$, and $\{342,210\}/2=$ $%
\{171,105\}$, correspondingly. Similarly, many sets of parameters $%
\left\{ \mu ,\nu \right\} $ solve the Eq. (\ref{totalS}), and we pick up the
one set which makes $\left\vert \mu \right\vert +\left\vert \nu \right\vert $
minimum too. The results are listed in the table III. The exact numerical
results confirm our mechanism as well. 
\begin{table}[tbp]
\caption{A numerical verification of the mechanism from saturated systems:  $%
\protect\mu \ll \min \left\{ \overline{\Delta N_{0in}},\overline{N_{in\text{ 
}ex}}\right\} ,$ $\protect\nu =0$.}$%
\begin{array}{|c|c|c|c|c|c|}
\hline
\left\{ N_{in},N_{out}\right\}  & \left\{ \overline{N_{0in}},\overline{%
N_{0out}}\right\}  & \left\{ \overline{\Delta N_{0in}},\overline{\Delta
N_{0out}}\right\}  & \left\{ S_{in},S_{out}\right\} /k & \Delta S/k & 
\left\{ \mu ,\nu \right\}  \\ \hline
\{231,45\} & \{411.69,74.01\} & \{26.85,8.40\} & \{14.09,5.69\} & -4.38 & 
\{0.28,0\} \\ \hline
\{210,66\} & \{372.90,111.02\} & \{25.12,11.09\} & \{13.42,7.15\} & -5.19 & 
\{0.13,0\} \\ \hline
\{171,105\} & \{301.14,180.89\} & \{21.76,15.42\} & \{12.08,9.31\} & -6.00 & 
\{0.41,0\} \\ \hline
\end{array}%
$%
\end{table}

\section{Conclusions and discussions}

That the particles in ground states do not have entropy is a fundamental
nature for boson systems. When two such systems merge, the number of
particles in ground states in the merged system is not a simple summation of
those in the original ones, introducing complexities and puzzles. To resolve
the puzzles, we propose a mechanism which holds with quite high accuracy. To
note that the mechanism is based on an anomalous particle-exchange between
excited-state particles that are fluctuating and variable in numbers, and is
verified by the ideal bosons on the surface of sphere. We wonder whether
there is a fundamental principle to interpret the mechanism, which is under
investigation.

\begin{acknowledgments}
This work is financially supported by Key Education Reform Projects in Hunan
Province under Grant No. HNJG-2023-0147.
\end{acknowledgments}

\end{document}